\documentclass[paper,prd,reprint,preprintnumbers,nofootinbib,notitlepage,showpacs,showkeys]{revtex4-1}
\usepackage{latexsym,graphicx,amssymb,amsmath,mathrsfs,color,enumerate}

\makeatletter
\renewcommand\appendix{\par
\setcounter{section}{0}%
\setcounter{subsection}{0}%
\gdef\thesection{\appendixname\space\@Alph\c@section}}
\long\def\unmarkedfootnote#1{{\long\def\@makefntext##1{##1}\footnotetext{#1}}}
\makeatother

\def \mate<#1|#2|#3>{\mbox{$\langle {#1}|\,{#2}\,|{#3}\rangle$}}

\newcommand{\br}{\mbox{\boldmath $r$}}
\newcommand{\bk}{\mbox{\boldmath $k$}}
\newcommand{\bzero}{\mbox{\boldmath $0$}}

\begin{document}

\title{Most Strange Dibaryon from Lattice QCD}

\author{
Shinya Gongyo$^{1}$,
Kenji~Sasaki$^{1,2}$,
Sinya~Aoki$^{1,2,3}$,
Takumi~Doi$^{1,4}$,
Tetsuo~Hatsuda$^{4,1}$,
Yoichi~Ikeda$^{1,5}$,
Takashi~Inoue$^{1,6}$,
Takumi Iritani$^{1}$,
Noriyoshi~Ishii$^{1,5}$,
Takaya~Miyamoto$^{1,2}$,
Hidekatsu~Nemura$^{1,5}$
\\ (HAL QCD Collaboration)\\
}
\affiliation{$^1${RIKEN Nishina Center, RIKEN,, Saitama 351-0198, Japan}}
\affiliation{$^2${Center for Gravitational Physics, Yukawa Institute for Theoretical Physics, Kyoto University, Kyoto 606-8502, Japan}}
\affiliation{$^3${Center for Computational Sciences, University of Tsukuba, Ibaraki 305-8571, Japan}}
\affiliation{$^4${RIKEN iTHEMS Program, RIKEN, Saitama 351-0198, Japan}}
\affiliation{$^5${Research Center for Nuclear Physics (RCNP), Osaka University, Osaka 567-0047, Japan}}
\affiliation{$^6${Nihon University, College of Bioresource Sciences, Kanagawa 252-0880, Japan}}

\begin{abstract}
The $\Omega\Omega$ system  in the $^1S_0$ channel (the most strange dibaryon) is studied on the basis of
 the (2+1)-flavor lattice QCD simulations with a large volume (8.1 fm)$^3$ and nearly 
  physical pion mass $m_{\pi}\simeq 146$ MeV at a lattice spacing $a\simeq 0.0846$ fm.  We show that lattice QCD data analysis by the HAL QCD method
  leads to the scattering length 
$a_0 = 4.6 (6)(^{+1.2}_{-0.5}) {\rm fm}$,
 the effective range $r_{\rm eff} = 1.27 (3)(^{+0.06}_{-0.03})  {\rm fm}$ and 
 the binding energy $B_{\Omega \Omega} = 1.6 (6) (^{+0.7}_{-0.6})   {\rm MeV}$.
  These results indicate that the $\Omega\Omega$ system has an overall attraction and is 
  located near the unitary regime.  Such a system can be best searched experimentally
  by the pair-momentum correlation in relativistic heavy-ion collisions.
\end{abstract}


\maketitle

\preprint{\today}
\preprint{RIKEN-QHP-320,YITP-17-91, RIKEN-iTHEMS-Report-17}

\noindent {\bf \em Introduction:}\ \ 
Dibaryon is defined as a baryon number $B$ = 2 system (equivalently  a 6-quark system)
  in quantum chromodynamics \cite{Mulders:1980vx,Oka:1988yq,Gal:2015rev}. 
  So far,   only one stable dibaryon, the deuteron, has been
  observed:  It is a loosely  bound system of the proton and the neutron in spin-triplet and
  isospin-singlet channel.  In recent years, there are renewed experimental interests in the 
  dibaryons due to exclusive measurements in hadron reactions \cite{Clement:2016vnl}
  as  well as the  direct measurement  in relativistic heavy-ion collisions \cite{Cho:2017dcy}. 
  Also from the theoretical side,  (2+1)-flavor lattice QCD simulations of the 6-quark system 
   in a large box with nearly physical quark masses became possible  recently \cite{Doi:2017cfx}. 
  The main aim of this Letter is to report the first result and physical implication of $\Omega\Omega$,  
  the strangeness $S = -6$  dibaryon (``most strange dibaryon"), in full QCD simulations
  with the lattice volume (8.1 fm)$^3$ and  the pion mass $m_{\pi} \simeq 146$ MeV at a lattice spacing $a\simeq 0.0846$ fm \cite{Ishikawa:2015rho}.  
  
  Before entering the detailed discussions  of our study, we first introduce the reason why
  such an exotic channel ($S=-6$) is of interest in QCD.
   Let us consider octet ${\bf 8}$ and decuplet ${\bf 10}$ baryons in the 
  flavor SU(3) classification.
  All the members of ${\bf 8}$ are stable under strong decay.
    This is why the forces between octet baryons in ${\bf 8} \otimes {\bf 8}$  are most relevant
   in the physics of hypernuclei and of neutron stars. 
   Also, the elusive $H$-dibaryon (a combination of
    $\Lambda\Lambda$, $N\Xi$ and $\Sigma \Sigma$) is in this representation \cite{Jaffe1977,Inoue:2010es,Beane:2010hg}
         and does not suffer from the Pauli exclusion principle in the flavor-SU(3) limit.
 
 On the other hand,   
  only $\Omega$ in ${\bf 10}$ is stable  under strong decay.
  Therefore,    in the ${\bf 8} \otimes {\bf 10}$    representation, the most
   promising candidate of  stable  dibaryon is $N\Omega$ \cite{Goldman:1987ma}.
    The Pauli exclusion principle does not work in this case too,
   so that  there is a possibility to have a bound state in the S-wave and total-spin 2 channel \cite{Etminan:2014tya}.
  Such a system  is indeed  studied by the two-particle momentum  correlation
     in high-energy heavy-ion collisions both theoretically and experimentally \cite{Morita:2016auo}. 
       
   In the decuplet-decuplet channnel, we have
 \begin{eqnarray}  
 {\rm   {\bf 10} \otimes {\bf 10} = ({\bf 28} \oplus {\bf 27})_{{sym.}} \oplus ({\bf 35} \oplus {\bf 10}^* )_{{anti\mathchar`-sym.}}  ,} \nonumber
 \end{eqnarray}    
 where "sym." and "anti-sym." stand for the flavor symmetry under the exchange of two baryons.
    Only possible stable state under strong decay is the $\Omega\Omega$
    system in the symmetric ${\bf 28}$ representation.  Again, the quark Pauli principle
    does not operate in this channel \cite{Oka:2000wj}.  Note that the celebrated  ABC resonance 
    ($\Delta\Delta$ in the spin-3 and isospin-0 channel)  \cite{Dyson:1964xwa,Clement:2016vnl}
    belongs to the anti-symmetric ${\bf 10}^*$ representation, 
    while  $\Delta\Delta$ in the spin-0 and isospin-3 channel
    is in the same multiplet with   $\Omega\Omega$.    
    
 The $\Omega\Omega$ interaction at low energies has been investigated so far
 by using  phenomenological quark models or by using lattice QCD simulations with heavy quark masses.
 Very recently, the chiral effective field theory has also been applied to the scattering of the $\Omega$ baryons \cite{Haidenbauer:2017sws}.
 In some quark models, strong attraction is reported  
 \cite{zhang:1997,zhang:2000}, while other models  show weak repulsion \cite{wang:1992,wang:1997}.
A  (2+1)-flavor lattice QCD study with $m_{\pi}\simeq 390$ MeV by using the finite volume  method  
shows  weak repulsion
\cite{Buchoff:2012prd}, while a study with $m_\pi \simeq 700$ MeV by using the HAL QCD method
shows  moderate attraction \cite{Yamada:2015cra}. Under such a controversial situation, it is most  
 important to carry out first-principles lattice QCD simulations in a large volume with  the pion mass close to the physical point.

\noindent {\bf \em HAL QCD method:}\ \ 
In the HAL QCD method \cite{Ishii:2006ec,Aoki:2010ptp,Ishii:2012plb,Aoki:2012tk}, 
 the observables such as the binding energy and phase shifts  are obtained 
 from the equal-time Nambu-Bethe-Salpeter (NBS) wave function $\psi(\br)$ 
and associated two-baryon irreducible kernel $U(\br,\br')$.
 The traditional finite volume method with the plateau fitting \cite{luscher:1991} is challenging in practice
 for $B=2$ systems in large volumes
 because of the difficulty in differentiating each scattering state
 \cite{Lepage:1990,Iritani:2016jie,Iritani:2017rlk,Aoki:2017byw}.  On the other hand,
  the time-dependent HAL QCD method  \cite{Ishii:2012plb} can avoid such a problem 
  since all the elastic scattering states are dictated by the same kernel $U(\br,\br')$
   and there is no need to identify each scattering state in a finite box.

The  equal-time NBS wave function $\psi(\br)$  has the property  that its asymptotic behavior at large distance
 reproduces the phase shifts, which can be proven from the  
 unitarity of $S$-matrix in quantum field theories  \cite{Ishii:2006ec}.
 Moreover, it is related to the following {\em reduced} four-point (4-pt) function,
 \begin{eqnarray}
 R(\br, t>0)  &=& \langle 0 \vert \Omega(\br,t)  \Omega(\bzero,t)    {\overline {\cal J}}(0) \vert 0 \rangle /e^{-2m_{_\Omega} t } \nonumber  \\ 
&=& \sum_n  a_n \psi_n(\br) e^{- (\delta W_n) t} + O(e^{- (\Delta E^*) t}). 
\label{eq:4pt} 
\end{eqnarray}
Here $  {\overline {\cal J}}(0)$ is a source operator  creating the $(B,S)=(2,-6)$ system  at Euclidean time 0,
and $a_n$ is the matrix element defined by  $\langle n \vert    {\overline {\cal J}}(0) \vert 0 \rangle$ 
with $ | n  \rangle $ representing the elastic states in a finite volume. The energy is represented as
$\delta W_n = 2\sqrt{m_{\Omega}^{2}+\bk_n^{2}} - 2m_{\Omega} $ with the $\Omega$ baryon mass $m_\Omega$ and the relative momentum $\bk _n$.
 Typical excitation energy of a {\em single} $ \Omega$-baryon is denoted by $\Delta E^*$, so that the last term in Eq.(\ref{eq:4pt}) 
is exponentially 
  suppressed as long as  $t \gg (\Delta E^*)^{-1} \sim \Lambda_{\rm QCD}^{-1}$  \cite{Ishii:2012plb}
 with  $\Lambda_{\rm QCD} \sim 200 - 300$ MeV  being the QCD scale parameter. 
A local interpolating operator for the $\Omega$ baryon
has a general form
\begin{eqnarray}
\Omega(x) &\equiv& \varepsilon^{abc} (s_{a}^{T}(x)C\gamma_{k}s_{b}(x))s_{c,\alpha}(x),
\label{eq:Omega}
\end{eqnarray}
with $a$, $b$ and $c$ being color indices, 
$\gamma_{k}$ being the Dirac matrix, $\alpha$ being the spinor  index, and 
$C\equiv\gamma_{4}\gamma_{2}$ being charge conjugation.  An appropriate spin projection
 is necessary from this operator to single out a particular spin state as mentioned later. 

The reduced 4-pt function $R$  has been shown to satisfy the following master equation \cite{Ishii:2012plb}, 
\begin{eqnarray}
(\frac{\triangledown^{2}}{m_{_\Omega}}-\frac{\partial}{\partial t}+\frac{1}{4m_{_\Omega}}\frac{\partial^{2}}{\partial t^{2}})R(\br, t)
= \int d\br' U(\br,{\br}')R({\br'}, t), \nonumber \\
\label{eq:tdep}  
\end{eqnarray}
which is valid as long as $t \gg (\Delta E^*)^{-1}$.
We emphasize that we do not need to isolate each scattering state with the energy $\delta W_n$, so that only the 
moderate values of $t $ up to 1.5-2 fm  are sufficient for a reliable extraction of the kernel $U$. This is a crucial difference from
the finite volume method which requires  $t \gg (\delta W_1)^{-1} \sim 10 $ fm (for the lattice volume as large as 8 fm)   to identify each $\delta W_n$. 
 (See a recent summary \cite{Aoki:2017byw}  and references therein.)
At low energies, one can make the derivative expansion with respect to the non-locality of  the kernel \cite{Aoki:2010ptp, Murano:2011nz}, 
$U(\br,\br')=V _0(r)\delta(\br-\br')+\sum_{n=1}V_{2n}(r)\nabla^{2n}\delta(\br-\br')$. 
Then, we introduce an ``effective" leading-order potential $V (r)$,
\begin{align}
V (r) =  R^{-1}(\br, t)  \left(\frac{\nabla^{2}}{m_{_\Omega}}-\frac{\partial}{\partial t}+\frac{1}{4m_{_\Omega}}\frac{\partial^{2}}{\partial t^{2}}\right)R(\br, t), \label{eq:Potential}
\end{align}
which  provides a good leading-order approximation of $U(\br,\br')$ to obtain physical observables at low energies,
as long as its $t$-dependence is sufficiently small.

\noindent {\bf \em  Interpolating operator:}\ \ 
The interpolating operator  $\Omega_{s_z}(x)$ 
 for  the $\Omega$ baryon with spin-$\frac{3}{2}$  and  the $z$-component $s_z=\pm \frac{3}{2}, \pm \frac{1}{2}$
 can be readily constructed by the appropriate spin projection of the upper two components of Eq.(\ref{eq:Omega}) as shown in \cite{Yamada:2015cra}.
The asymptotic $\Omega\Omega$ system can now be characterized by $^{2s+1}L_J$ with
the total spin $(s)$, 
the orbital angular momentum $(L)$ and the total angular momentum $(J)$. 
We consider $L=0$  where maximum attraction is expected at low energies.
 Then, the Fermi statistics leads  $s$ to be even (either $s$=0 or  2). 
  Here we consider an  $s=0$ system with the interpolating operator
\begin{eqnarray}
\label{eq:OO_00}
[\Omega\Omega]_{0} =
\frac{1}{2} 
 \left(
 \Omega_{\frac{3}{2}}\Omega_{-\frac{3}{2}}
-\Omega_{\frac{1}{2}}\Omega_{-\frac{1}{2}}  
 +\Omega_{-\frac{1}{2}}\Omega_{\frac{1}{2}}
-\Omega_{-\frac{3}{2}}\Omega_{\frac{3}{2}}
\right) .  \nonumber \\
\end{eqnarray}

For ${\overline {\cal J}}(0)$, we use the wall-type quark source with the $s=0$ projection given above.
 With this source, the total momentum of the system is automatically zero. Also,
it has good overlap with  the scattering states where  $|{\br}|$ in Eq.(\ref{eq:tdep})
 is larger than the typical baryon size. 
To extract the  $L=0$ and $s=0$  state at $t$ on the lattice,  we employ 
 Eq.(\ref{eq:OO_00}) for the sink operator together with the projection to the 
 $A_1$ representation of the cubic-group, 
\begin{equation}
P^{(A_1)} R({\br}, t) =  \frac{1}{24}\sum_{i=1}^{24} R({\cal R}_{i}[{\br}],t),
\end{equation}
where ${\cal R}_{i}$ is an element of the cubic group acting on the relative distance $\br$.

Note here that $R({\br}, t)$ and $U({\br},{\br'})$ depend on the choice of interpolating operators,
 while observables calculated from these quantities are independent of the choice thanks to the 
 Nishijima-Zimmermann-Haag theorem \cite{Aoki:2010ptp}.

\noindent {\bf \em Lattice Setup:}\ \ 
By using the 11PFlops supercomputer K at RIKEN Center for Computational Science,
  $(2+1)$-flavor gauge configurations on the $96^4$ lattice are generated 
with the Iwasaki gauge action at $\beta = 1.82$ and 
nonperturbatively ${\cal O}(a)$-improved Wilson quark action with stout smearing.
The lattice spacing is  $a \simeq 0.0846$ fm ($a^{-1} \simeq 2.333$ GeV)  \cite{Ishikawa:2015rho}
 and  the pion mass, the kaon mass and the nucleon masses are  
  $m_\pi \simeq 146$ MeV, $m_K \simeq 525$ MeV and $m_N \simeq 964$ MeV, respectively.
  (These masses are  higher than the physical values by about 8\%, 6 \% and 3 \%, respectively,
  due to slightly larger quark masses at the simulation point.)
   The lattice size, $La \simeq 8.1$ fm, is sufficiently large
to accommodate two baryons in a box.

We employ the wall quark source with the Coulomb gauge fixing, 
and the periodic (Dirichlet) boundary condition is used for spatial (temporal) directions.
Forward and backward propagations are averaged to reduce the statistical fluctuations.
We pick 1 configuration per each 5 trajectories,
and make use of the rotation symmetry and the translational invariance for the source position to increase the statistics.
The total statistics in this Letter amounts to
400 configurations $\times$ 2 (forward/backward) $\times$ 4 rotations $\times$ 48 source positions.
The quark propagators are obtained by the domain-decomposed solver \cite{Boku:2012zi,Terai:2013, Nakamura:2011my,Osaki:2010vj}
and the correlation functions are calculated using the unified contraction
algorithm \cite{Doi:2012xd}.

The $\Omega$-baryon mass extracted from the 
 effective mass $m_{\rm eff}(t) \equiv \ln G(t)/G(t+a)$ 
 with  $G(t)$ being the baryonic two-point function 
  is  $m_{_\Omega} = 1712 \pm 1 $ MeV  (from the plateau
  in $t/a=17-22$) and 
 $m_{_\Omega}= 1713 \pm 1 $ MeV (from $t/a =18-25$) with the statistical errors.
 These numbers are  about  2\% higher than the physical value, 1672 MeV.
We take the former number in the following analysis.

\begin{figure}[tb]
\begin{center}
\includegraphics[scale=0.66]{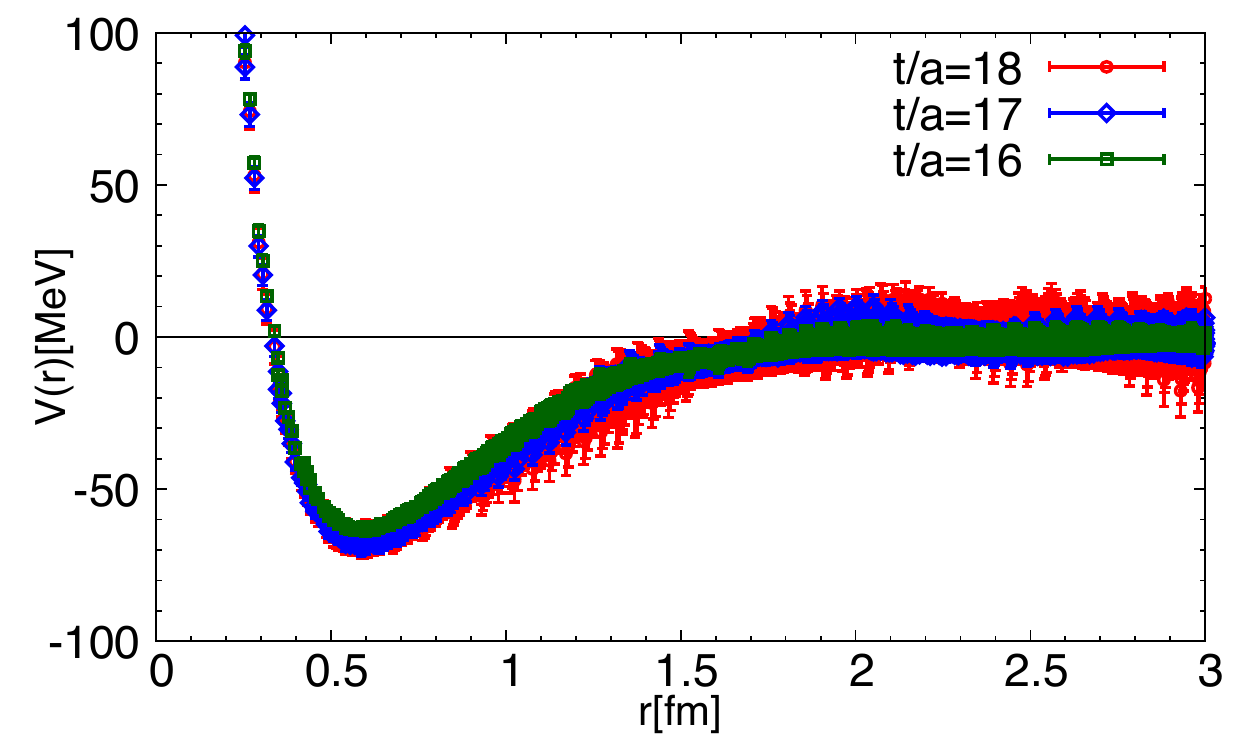}
\caption{The $\Omega \Omega$ potential $V(r)$ in the $^1S_0$ channel at 
Euclidean time $t/a=16,17,$ and 18.}
\label{fig:Potential}
\end{center}
\end{figure}

\noindent{\bf \em Numerical results:} \ \ 
The $^1S_0$ potential $V(r)$ obtained from   Eq.(\ref{eq:Potential})
 with the lattice measurement of $R(\br, t)$ 
 is shown in Fig. \ref{fig:Potential} for $t/a=16, 17,$ and 18.
 Here  Laplacian and the time-derivative in Eq.(\ref{eq:Potential}) 
  are approximated by the central (symmetric) difference.
 The statistical errors  for $V(r)$  at each  $r$ are estimated 
 by the jackknife method with the bin size of 40 configurations.
A comparison with the bin size of 20 configurations shows that the bin size dependence is small.
     The particular region $t/a=17 \pm 1$ in Fig. \ref{fig:Potential} is chosen to suppress
 contamination from excited states in the single $\Omega$ propagator at smaller $t$ and simultaneously to
  avoid  large statistical errors at larger $t$.  We observe that the potentials at $t/a= 16, 17,$ and 18  are
   nearly identical  within statistical errors as expected from the  time-dependent HAL QCD method \cite{Ishii:2012plb}. 
      
  The $\Omega\Omega$ potential $V(r)$ has qualitative features similar to the central potential of the  nucleon-nucleon ($NN$) interaction,
   i.e., the short range repulsion and the intermediate range 
  attraction \cite{Doi:2017cfx}.  There are, however, two quantitative differences:  (i)
   the short range  repulsion is much weaker in the $\Omega\Omega$ case
    possibly due to the absence of quark Pauli exclusion effect, and (ii) the  attractive part is much 
    short-ranged due to the absence of  pion exchanges. 
    
   For  the purpose of converting  the potential
   to the  physical observables such as the scattering phase shifts and the binding energy,
   we  fit $V(r)$  in  Fig. \ref{fig:Potential} in the range $r =0-6\  {\rm fm}$ 
   by  three Gaussians,   $V_{\rm fit}(r) = \sum_{j=1,2,3} c_j \exp(-(r/d_j)^2)$. 
    For example, the uncorrelated fit  in the case of  $t/a=17$ gives the following parameters:
    $(c_1, c_2, c_3)= (914(52), 305(44), -112(13))$ in MeV
    and  $(d_1, d_2, d_3)=0.143(5), 0.305(29), 0.949(58))$ in fm  with
 $\chi^2/{\rm d.o.f.} \sim 1.3$.    Another functional form such as two Gaussians + (Yukawa function)$^2$
     provides  equally well fit, and the results are not affected  within 
     errors.  The finite volume effect on the potential is expected to be small due to the large lattice size. 
     The naive order estimate of the finite $a$ effect for the physical observables is also small ($(\Lambda a)^2 \le 1$ \%) thanks to 
     the non-perturbative $O(a)$ improvement, but an explicit confirmation would be desirable in the future.      
         
\begin{figure}[t]
\begin{center}
\includegraphics[scale=0.7]{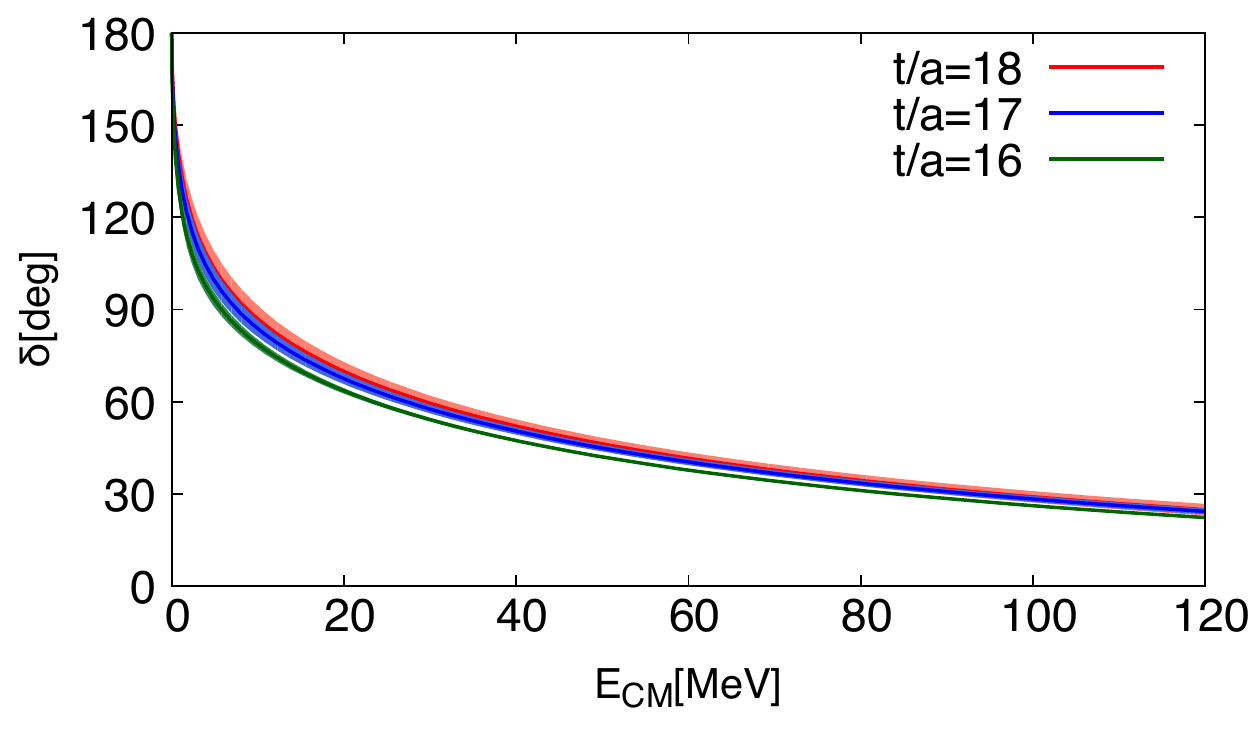}
\caption{The $\Omega \Omega$ phase shift $\delta(k)$ in the $^1S_0$ channel
 for $t/a=16, 17$ and 18  as a  function of the center of mass kinetic energy
 $E_{_{\rm CM}}=2 \sqrt{k^2+m_{_\Omega}^2 } - 2m_{_\Omega}$. }
\label{fig:Phase-shift}
\end{center}
\end{figure}

   The $\Omega \Omega$ scattering phase shifts $\delta(k)$ in the 
   $^1S_0$ channel obtained from $V_{\rm fit}(r)$ 
  are shown in Fig.\ref{fig:Phase-shift}  for $t/a=16, 17,$ and $18$  as a function of the 
  kinetic energy in the center of mass frame, 
  $E_{_{\rm CM}}=2 \sqrt{k^2+m_{_\Omega}^2 } - 2m_{_\Omega}$. 
  The error bands
 reflect the statistical uncertainty of the potential in Fig. \ref{fig:Potential}. 
     All three cases show that $\delta(0)$ starts from 180$^{\circ}$, which  
   indicates the existence of a  bound $\Omega\Omega$ system.

\begin{figure}[tb]
\begin{center}
\includegraphics[scale=0.70]{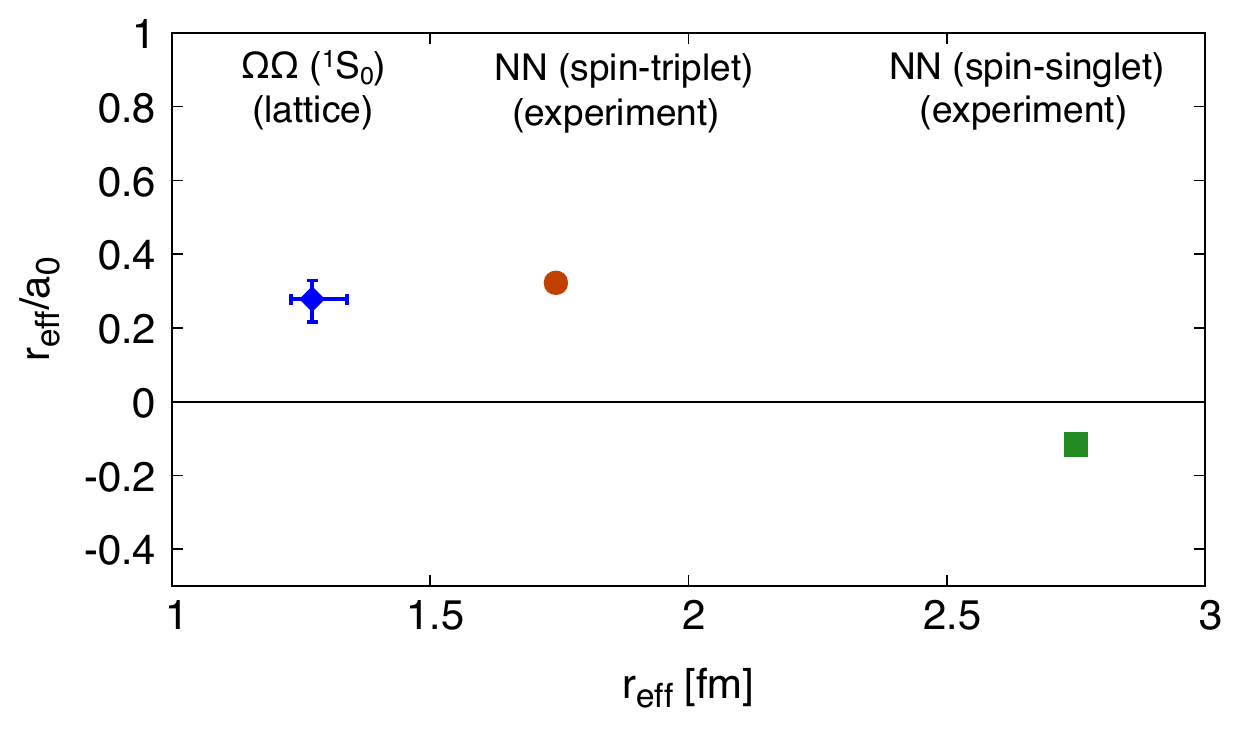}
\caption{The dimensionless  ratio of the effective range $r_{\rm eff}$ and the  scattering length $a_0$ 
 as a function of $r_{\rm eff}$ for the $\Omega\Omega$ system in the $^1S_0$ channel as well as for the 
  spin-triplet  $NN$ system (the deuteron channel) and  for the spin-singlet $NN$ system (the neutron-neutron channel).
  The error bar for $\Omega\Omega$ is the quadrature of the statistical and systematic errors in Eqs. (\ref{eq:a0}) and (\ref{eq:r0}).}
\label{fig:Scattering}
\end{center}
\end{figure}

 The scattering length $a_0$  and the effective range $r_{\rm eff}$ in the $^1S_0$ channel 
 is extracted from $\delta(k)$ through the effective range expansion (ERE), 
   $k \cot \delta(k) = - \frac{1}{a_0} + \frac{1}{2} r_{\rm eff} k^2 + \cdots$, with
   the sign convention of  nuclear and atomic physics:
\begin{eqnarray}
a_0^{(\Omega \Omega)} &=& 4.6 (6)(^{+1.2}_{-0.5}) \ \  \ {\rm fm}, \label{eq:a0} \\
r_{\rm eff}^{(\Omega \Omega)} &=& 1.27 (3)(^{+0.06}_{-0.03}) \ \ {\rm fm}.
 \label{eq:r0}
\end{eqnarray}
 The central values and the statistical errors in the  first parentheses are  extracted from $\delta(k)$ 
 at  $t/a=17$, and  the  systematic errors  in the second parentheses 
 are estimated from the results at $t/a=16$ and 18. The origin of this systematic error is
  the truncation of the higher-derivatives of the non-local potential as well as
  the contaminations from inelastic states.
  To get a feel for  the magnitude of these values, we recapitulate here the experimental values of
  $a_0$ and $r_{\rm eff}$  in the $NN$ systems; 
 $(a_0, r_{\rm eff})_{\rm spin\mathchar`- triplet}=(5.4112(15)  {\rm fm}, 1.7436(19) {\rm fm})$ 
 and  $(a_0, r_{\rm eff})_{\rm spin\mathchar`-singlet}=(-23.7148(43)  {\rm fm}, 2.750(18)  {\rm fm})$ \cite{Hackenburg:2006qd}.
  There exists no symmetry reason that the  scattering parameters in the $NN$ systems and those in the 
  $\Omega\Omega$ system should be similar. Nevertheless,  
  it is remarkable that they are all close to  the unitary region where  $r_{\rm eff}/a_0 $ is substantially smaller than 1
   as shown in Fig.\ref{fig:Scattering}.

\begin{figure}[tb]
\begin{center}
\includegraphics[scale=0.70]{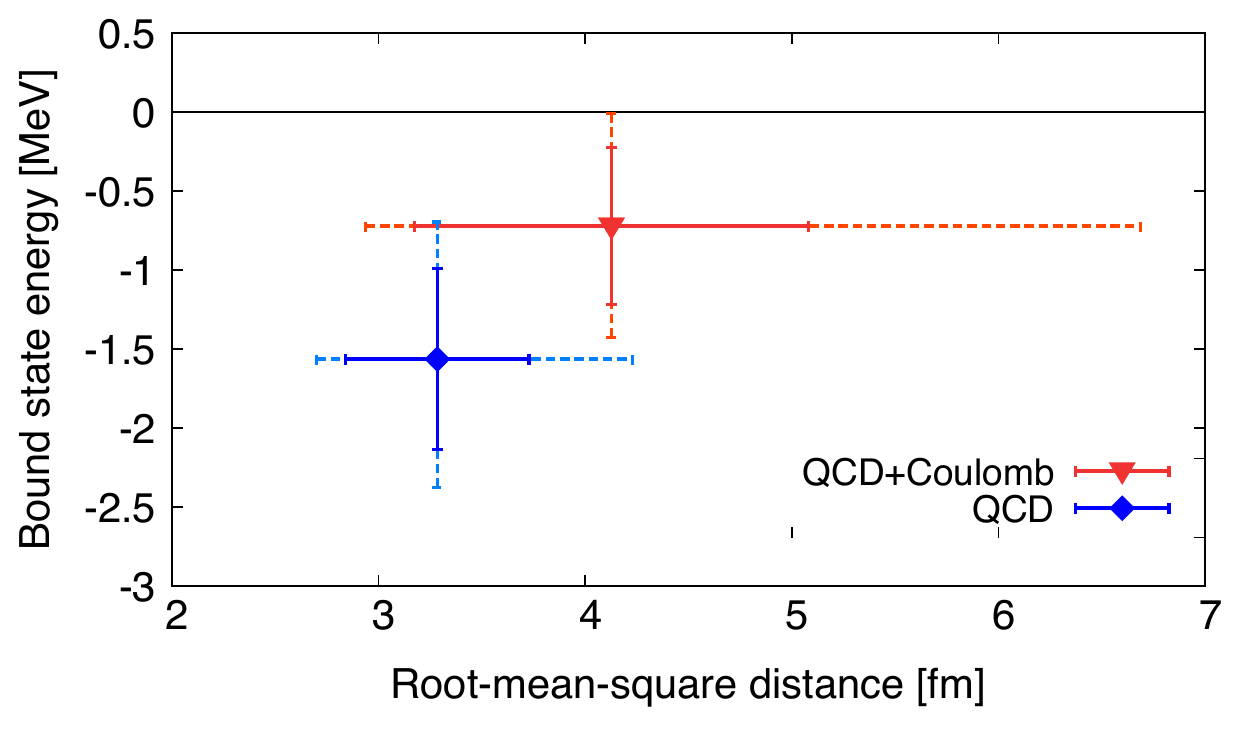}
\caption{Bound state energy of the $\Omega\Omega$ system and the root-mean-square distance between $\Omega$s
 obtained from the potential.  Filled diamond (triangle) corresponds to the result at $t/a=17$ without (with) the Coulomb 
  repulsion.  The statistical errors are shown by the solid lines, while the systematic errors estimated from the 
  difference between the data at $t/a=17$ and those
 at $t/a=16, 18$ are shown by the dashed lines.  }
\label{fig:Binding}
\end{center}
\end{figure}

 Shown in Fig.\ref{fig:Binding} are  the bound state energy given by the opposite sign of the binding energy, $-B_{\Omega\Omega}$, and
    the root-mean-square distance ($\sqrt{\langle r^2 \rangle}$) of the $\Omega\Omega$ bound state obtained from the potential.
    The blue diamond is taken from the data at $t/a=17$ without the Coulomb repulsion.  The blue solid and dashed lines are the 
     statistical error at $t/a=17$ and the systematic error estimated from the data at $t/a=17\pm 1$, respectively:
 \begin{eqnarray}
 B_{\Omega \Omega}^{\rm (QCD)} = 1.6 (6) (^{+0.7}_{-0.6}) \ \  {\rm MeV}.
\end{eqnarray} 
  As an alternative estimate, the truncation error of the derivative expansion on the binding energy is 
  evaluated perturbatively, and is found to be less than $20\%$ even if
  the magnitude of dimensionless next-to-leading order potential is the same order
  as that of the effective leading-order potential.
 It is an important future subject to determine higher-order potentials explicitly
   by using the method of multiple quark sources \cite{Iritani:2017wvu}.
The binding energy is consistent with the value
   obtained from the  general formula for loosely bound states \cite{Naidon:2016dpf} with 
  (\ref{eq:a0})  and   (\ref{eq:r0});  
  $B_{\Omega\Omega}= \frac{1}{m_{\Omega} r_{\rm eff}^2} \left( 1-\sqrt{1-\frac{2r_{\rm eff}}{a_0}} \right)^2 \simeq 1.5$ MeV.  
 Associated with this small binding energy,  $\sqrt{\langle r^2 \rangle}$ is as large as 3-4 fm 
 which is consistent with the expectation, $\sqrt{\langle r^2 \rangle} \sim a_0$, for
  loosely bound states. The Coulomb repulsion can be evaluated by adding $\alpha/r$ with $\alpha = e^2/4\pi$ to the potential obtained from lattice QCD, i.e. $V^{(\mathrm{QCD}+ \mathrm{Coulomb})} \equiv V^{(\mathrm{QCD})} + \alpha/r$. This reduces the above binding energy by 
  a factor of two,  $B_{\Omega \Omega}^{\rm (QCD+Coulomb)} = 0.7 (5) (5) \  {\rm MeV}$ 
    as shown in Fig.\ref{fig:Binding} by the red triangle. 
 
  It is in order here to remark that there are three energy scales in the present problem:
    $2m_{_\Omega} \simeq $ 3400 MeV $\gg$ $ \vert V(r\simeq 0.5 {\rm fm})\vert  \sim$  50 MeV $\gg$ $B_{\Omega\Omega} \sim 1$ MeV.
   Since only the relative difference between the interacting and non-interacting two-$\Omega$ systems 
   matters, the absolute magnitude of the uncertainty of $2m_{_\Omega}$
    is not reflected directly in $V(r)$.  This is why we could extract $V(r)$ rather accurately as shown
    in Fig.\ref{fig:Potential} despite of the large scale difference between  $2m_{_\Omega}$ and $V(r)$.
     Then the small binding energy $B$  as well as  the large scattering length $a_0$ are the natural consequence of
     the large cancellation between the long-range attraction and the short-range repulsion of $V(r)$, a situation common
     in nuclear and atomic physics.  Although $V(r)$ is not a direct observable,  it provides
     an important intermediate step to link the QCD scale (GeV) to nuclear physics scale (MeV), since it is difficult to
     measure the binding energy directly from lattice QCD using the finite volume method for large lattice volumes 
     and physical quark masses (see the critical review \cite{Aoki:2017byw}).
    
   Finally let us estimate the effect of slightly heavy quark masses in our simulation.
   First of  all, the attractive part of the $\Omega\Omega$ potential would be slightly
   long-ranged at the physical point due to the kaon mass, $m_K^{(\rm present)}\simeq 525$ MeV $\rightarrow$
    $m_K^{(\rm phys.)} \simeq 495$ MeV.  On the other hand, the effect of the $\Omega$ mass,
    $m_{_\Omega}^{(\rm present)}\simeq 1712$ MeV $\rightarrow$
    $m_{_\Omega}^{(\rm phys.)}  \simeq 1672$ MeV,  would lead to less-binding due to the larger kinetic energy.
    Therefore, conservative estimate is obtained by keeping the same $V(r)$ in Fig.\ref{fig:Potential}
    and  to adopt     $m_{_\Omega}^{(\rm phys.)} $ to calculate the phase shifts and the binding energy.
    This results in  $(a_0, r_{\rm eff})= (4.9(8) {\rm fm}, 1.29(3){\rm fm})$ 
    and $(B_{\Omega \Omega}^{\rm (QCD)},B_{\Omega \Omega}^{\rm (QCD+Coulomb)} )=(1.3(5) {\rm MeV}, 0.5(5) {\rm MeV})$
    for the potential at $t/a=17$.  These numbers are well within errors of the results of the present simulation
    shown in Figs. \ref{fig:Scattering} and \ref{fig:Binding}.

\noindent{\bf \em Summary and Discussions:} \ \ 
In this Letter, we presented a first realistic calculation on the most strange dibaryon, $\Omega\Omega$, in the $^1S_0$ channel
 on the basis of the (2+1)-flavor lattice QCD simulations with a large volume and nearly physical quark masses.
 The scattering length, effective range and the binding energy obtained by the HAL QCD method strongly indicate
  that the $\Omega\Omega$ system in the $^1S_0$ channel has an overall attraction and is located in the vicinity of the unitary regime.
  From the phenomenological point of view, such a system can be best searched by the measurement of 
  pair-momentum correlation $C(Q)$ with $Q$ being
the   relative momentum between two baryons produced in relativistic heavy-ion collisions  \cite{Cho:2017dcy}.
   Experimentally,    each $\Omega$ can be identified
 through a successive weak decay, $\Omega^- \rightarrow \Lambda + K^-  \rightarrow p+\pi^- + K^-$.   
    Note that  a large scattering length (not the existence of a bound state)
  is the important element for $C(Q)$  to have characteristic enhancement  at small relative momentum $Q$.
 Moreover,  the effect of the Coulomb interaction can be   effectively eliminated by taking a ratio of 
 $C(Q)$ between small and large collision systems   \cite{Morita:2016auo}.

   We are currently underway to increase the statistics of the lattice data with the same lattice setup.  
   These results together with the detailed examination of the spectrum analysis in a
   finite lattice volume and the effective range expansion will be reported.

\begin{acknowledgments}

We thank members of PACS Collaboration for the gauge configuration generation.
The lattice QCD calculations have been performed on the K computer at RIKEN, AICS
(hp120281, hp130023, hp140209, hp150223, hp150262, hp160211, hp170230),
HOKUSAI FX100 computer at RIKEN, Wako (G15023, G16030, G17002)
and HA-PACS at University of Tsukuba (14a-20, 15a-30).
We thank ILDG/JLDG~\cite{conf:ildg/jldg}
which serves as an essential infrastructure in this study.
We thank the authors of cuLGT code~\cite{Schrock:2012fj} used for the gauge fixing.
This study is supported in part by Grant-in-Aid for Scientific Research on Innovative Areas(No.2004:20105001, 20105003) 
and for Scientific Research(Nos. 25800170, 26400281), 
SPIRE (Strategic Program for Innovative REsearch), MEXT Grant-in-Aid for Scientific Research (JP15K17667, JP16H03978, JP16K05340),
"Priority Issue on Post-K computer" (Elucidation of the Fundamental Laws and Evolution of the Universe),
and by Joint Institute for Computational Fundamental Science (JICFuS). S.G. is supported by the Special Postdoctoral Researchers Program of RIKEN.
T.D. and T.H. are partially supported by RIKEN iTHES Project and iTHEMS Program.
T.H. is  grateful to the Aspen Center for Physics, supported in part by NSF Grants PHY1066292 and PHY1607611.
The authors thank C. M. Ko for drawing out attention to the $\Omega\Omega$ system, and K. Yazaki for fruitful discussions on the short range part of baryon-baryon interactions, and Y. Namekawa for his careful reading of the manuscript.	

\end{acknowledgments}


\end{document}